\documentclass[letter]{aa}
\usepackage{graphicx}
\usepackage{txfonts}
\pdfmapfile{+local_txfonts.map}
\pdfmapfile{+pdftex35.map}
\pdfmapfile{+cm.map}
\usepackage{amsmath}
\usepackage{xcolor}
\usepackage{placeins}
\usepackage{balance}
\usepackage{accsupp}
\usepackage{hyperref}
\hypersetup{colorlinks=true,citecolor=blue,linkcolor=blue,urlcolor=blue}
\newcommand{\scobj}{2025~SC$_{79}$}
\makeatletter
\DeclareRobustCommand{\saltillo}{\ifx\f@family\sfdefault\includegraphics[height=1em]{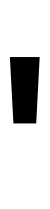}\else\includegraphics[height=1em]{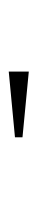}\fi}
\makeatother
\DeclareRobustCommand{\ayloname}{\texorpdfstring{\BeginAccSupp{method=hex,unicode,ActualText=A78C00410079006C00F3A78C0063006800610078006E0069006D}\saltillo Ayl\'o\saltillo chaxnim\EndAccSupp{}}{ꞌAylóꞌchaxnim}}

\newlength{\SCsavedfloatsep}
\newlength{\SCsavedtextfloatsep}
\begin{document}
\title{VLT and Keck recovery and characterisation of \scobj: linking \ayloname\ and Atira asteroids}
\author{Bryce T. Bolin\inst{1}
\and Leo Y. Alcorn\inst{2}
\and Davide Farnocchia\inst{3}
\and Marco Delbo\inst{4}
\and Martina Baratella\inst{5}
\and Henri M. J. Boffin\inst{6}
\and Max Brodheim\inst{2}
\and Alexa Gordon\inst{7}
\and Lordrick Kahinga\inst{8}
\and Michael Lundquist\inst{2}
\and Rita Morris\inst{2}
\and J. Xavier Prochaska\inst{8}}
\institute{Eureka Scientific, Inc., 2452 Delmer Street, Suite 100, Oakland, CA 94602-3017, USA
\email{bolin.astro@gmail.com}
\and W. M. Keck Observatory, 65-1120 Mamalahoa Highway, Kamuela, HI 96743, USA
\and Jet Propulsion Laboratory, California Institute of Technology, Pasadena, CA, USA
\and Laboratoire Lagrange, CNRS, Observatoire de la C\^ote d'Azur, Boulevard de l'Observatoire, CS 34229, 06304 Nice Cedex 4, France
\and European Southern Observatory, Santiago, Chile
\and European Southern Observatory, Garching, Germany
\and Department of Astronomy, California Institute of Technology, MC 249-17, 1200 East California Boulevard, Pasadena, CA 91125, USA
\and Department of Astronomy and Astrophysics, University of California, Santa Cruz, 1156 High Street, Santa Cruz, CA 95064, USA}
\titlerunning{2025 SC79: linking \ayloname\ and Atira asteroids}
\authorrunning{B. T. Bolin et al.}
\date{}
\abstract
{Asteroids near the boundary between the Atira and \ayloname\ classes can trace dynamical links between the two populations, but their small solar elongations make recovery and physical characterisation difficult.}
{We recover \scobj, measure its optical colours, and investigate its likely source region and transitions between these populations.}
{We combine Keck/LRIS astrometry and VLT/FORS2 four-filter photometry with Bus--DeMeo taxonomic comparisons, published source-model probabilities, and integrations of 10\,000 clones drawn from the covariance of JPL solution~\#2 and followed for 5~Myr in each direction.}
{The SDSS-equivalent colours are $g-r=0.83_{-0.13}^{+0.21}$, $r-i=0.176_{-0.054}^{+0.052}$ and $i-z=-0.50_{-0.12}^{+0.11}$~mag. The V-type class mean gives the closest colour match. Adopting the V-type median albedo $p_V=0.362$, we infer a diameter of 0.38~km from the catalogue absolute magnitude $H_V=18.83$~mag. The source model assigns a combined probability of 95.4\% to delivery channels associated with the inner Main Belt. The ensemble-mean \ayloname/Atira residence fractions are 30.2\%/69.8\% over the preceding 100~kyr and 3.6\%/96.0\% over the following 100~kyr; the remaining 0.4\% of future ensemble time follows collision removal. Clones spend an average of 231~kyr in the \ayloname\ class over the preceding 500~kyr and 120~kyr over the following 500~kyr. In the 5-Myr forward integrations, 14.56\% of clones collide with Venus, 3.77\% with Mercury and 0.22\% with Earth.}
{The V-like colours and source-model probabilities favour an origin in the inner Main Belt for \scobj. Its clones move between the \ayloname\ and Atira populations.}
\maketitle

\section{Introduction}\label{sec:intro}
Asteroids with aphelion distances $Q<0.983$~au remain interior to Earth's perihelion. Within this population we distinguish Atiras from \ayloname\ asteroids, whose aphelia lie inside Venus's perihelion ($Q<0.718$~au). \citet{Rondon2026Atiras} counted 37 Atiras and one \ayloname\ asteroid in their sample of known inner-Earth objects. These objects appear at small solar elongations. In ground-based twilight surveys, enhanced sky brightness and high airmass reduce sensitivity, and the survey footprint and cadence limit coverage \citep{Bolin2022Aylo,Bolin2025Twilight}. (163693) Atira has colours consistent with a dark surface \citep{Bolin2026Atira}, while (594913) \ayloname\ is red and reaches aphelion at only 0.65~au \citep{Bolin2022Aylo}.

S.~S.~Sheppard discovered \scobj\ at $r\simeq21$~mag on 2025 September 27 with the Dark Energy Camera on the 4-m Blanco telescope at Cerro Tololo \citep[MPEC~2025-T164]{MPC2025SC79}. Its aphelion of 0.719~au lies just beyond Venus's perihelion. We combine the Keck recovery and VLT follow-up with orbital integrations to characterise its surface and investigate dynamical links between the two classes. We compare its orbit with Main Belt delivery models to constrain its likely source region \citep{Granvik2018NEO,Nesvorny2023NEOMOD}.

\section{Keck and VLT observations}\label{sec:recovery}
\begin{figure*}[!t]
\centering\includegraphics[width=180mm]{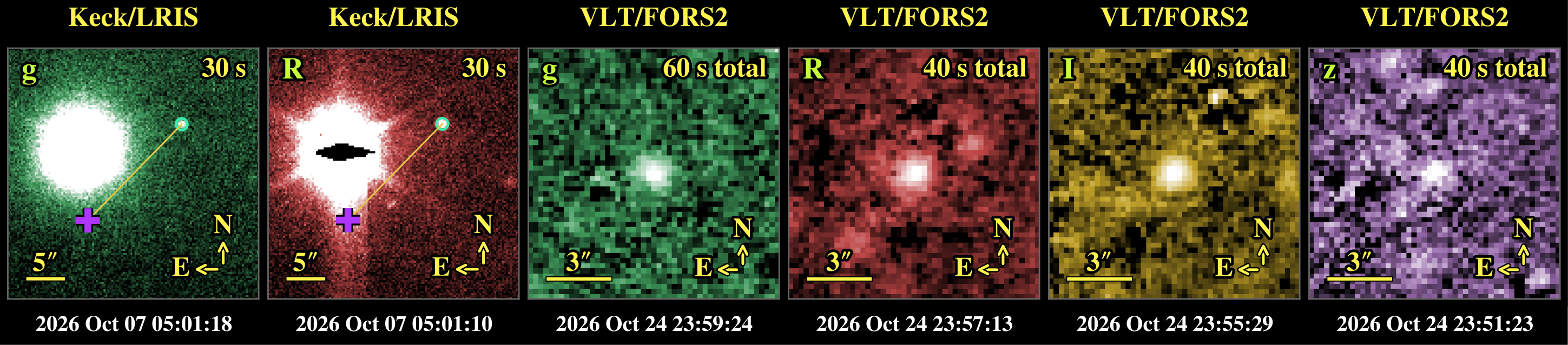}
\caption{Keck/LRIS recovery images in $g$ and $R$ (2026 October 7) and target-aligned VLT/FORS2 stacks in $g,R,I,z$ (October 24--25). Labels give the instrument, filter and exposure time. The UTC midpoint appears beneath each stamp. For each FORS2 stack we give the arithmetic mean of the constituent exposure midpoints. In each Keck image a yellow line joins the measured position (green circle) to the historical pre-recovery JPL~\#1 prediction (purple cross), reconstructed from the archived 19-arcsec offset. The Keck images use independent linear stretches. The FORS2 panels are 12.25~arcsec across and use independent asinh stretches. We subtracted neighbours, normalised by exposure time and averaged 3, 2, 2 and 2 frames, respectively, on a 0.25-arcsec grid. Each panel shows a scale bar at lower left and north and east at lower right.}\label{fig:recovery}\label{fig:fors2}
\end{figure*}

The discovery apparition spanned 2025 September 27--October 8. JPL solution~\#1 used 19 observations from 2025 September 27 through October 7 to predict the position at the next apparition. On 2026 October 7, we recovered \scobj\ with four 30-s Keck~I/LRIS exposures, two in $g$ and two in $R$, under programme 2026B\_K450 (PI: L.~Y.~Alcorn). The Keck recovery and FORS2 follow-up extended the observed arc to 374 and 392~d, respectively. The 19-arcsec offset to the north-west (Fig.~\ref{fig:recovery}) lies inside the formal $1\sigma$ prediction ellipse, with semiaxes of approximately 382 and 24~arcsec.
\begin{figure*}[!t]
\centering\includegraphics[width=165mm]{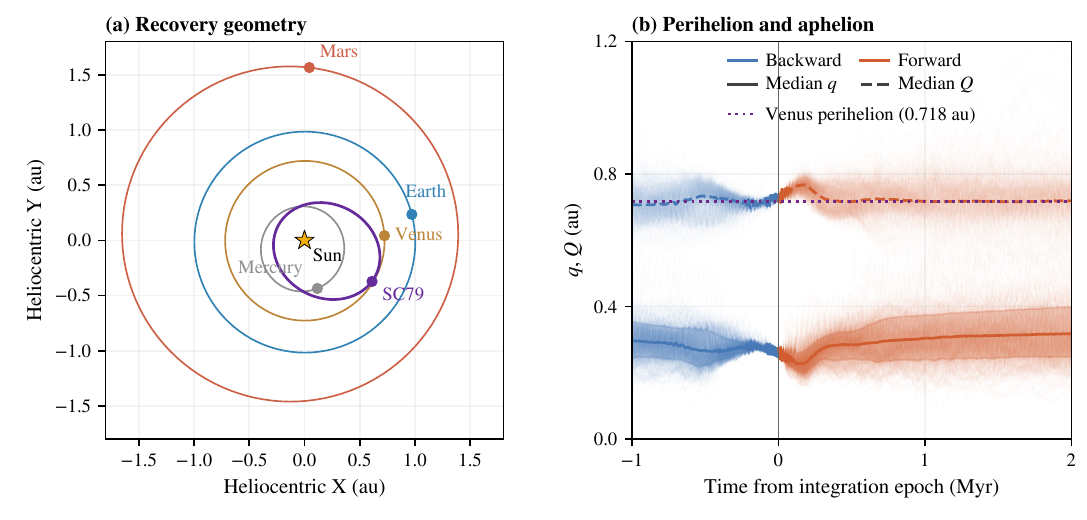}
\caption{Orbital geometry and evolution of \scobj. (a) Heliocentric orbits projected onto the J2000 ecliptic, viewed from the north, at the midpoint of the first Keck exposure (2026 October 7, 05:00:04.370~UTC). Symbols mark positions. Ellipses are osculating orbits constructed from JPL Horizons state vectors. For \scobj, solution~\#2 (Table~\ref{tab:orbit}) is propagated to the recovery epoch. (b) Perihelion $q$ and aphelion $Q$ for 400 JPL~\#2 clones sampled every 1000~yr before (blue) and after (orange) JD~2461174.5~TDB. Solid and dashed curves give median $q$ and $Q$, respectively, over the surviving ensemble. Shading spans the 16th--84th percentiles of $q$. The purple dotted line marks Venus's adopted perihelion distance (0.718~au).}\label{fig:orbit}\label{fig:qQ}
\end{figure*}

\begin{figure*}[!t]
\centering\includegraphics[width=180mm]{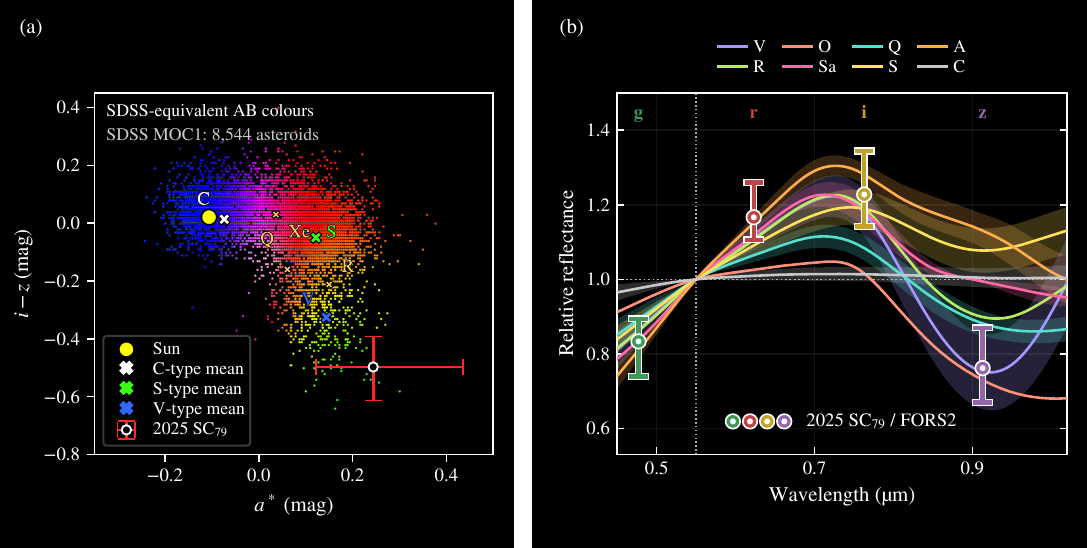}
\caption{FORS2 colour and reflectance diagnostics. (a) SDSS-equivalent $i-z$ versus $a^*$. Red error bars mark \scobj, coloured points represent 8544 SDSS asteroids, crosses mark selected Bus--DeMeo templates and the yellow circle marks the Sun. (b) Filter-coloured circles show the composite FORS2 reflectances of \scobj\ at nominal $g,r,i,z$ wavelengths, normalised by the mean of the solar-corrected $g$ and $r$ reflectances as a proxy for the reflectance at 0.55~$\mu$m. Filter-coloured error bars outlined in white show the propagated asymmetric colour uncertainties. The composite combines the exposure selections in Table~\ref{tab:colours}. Bus--DeMeo curves \citep{DeMeo2009} are normalised at 0.55~$\mu$m and interpolated with cubic splines. Shading shows the class dispersion.}\label{fig:astar}\label{fig:spectrum}
\end{figure*}
FORS2 obtained twenty exposures of approximately 20~s, five in each of $g_{\rm HIGH},R_{\rm SPECIAL+76},I_{\rm BESS+77},z_{\rm GUNN}$, on October 24--25 under programme 116.2AP3.002 (PI: B.~T.~Bolin). Airmass ranged from 1.87 to 2.24 and the stellar full width at half maximum (FWHM) from 0.81 to 1.76~arcsec. The heliocentric distance was $r_{\rm h}=0.697$~au, the distance from the observer was $\Delta=0.667$~au and the phase angle was $\alpha=93.6^\circ$. We interleaved five $g,R,I,z$ cycles over 27.35~min, revisiting each filter approximately every 5.7~min; photometry uses individual exposures.
\section{Astrometry and photometry}\label{sec:measurements}
The Gaia DR3 plate solutions give adopted Keck uncertainties of 0.10--0.11~arcsec per coordinate for three positions and 0.31~arcsec for the first $R$ frame. Appendix~\ref{app:newastrometry} lists four Keck $g/R$ positions and four FORS2 $R$ positions, excluding one blended FORS2 measurement. We use JPL~\#2, including the T16 recovery, for the orbital elements (Table~\ref{tab:orbit}), geometry, residuals and clone initial conditions. The tabulated aphelion places \scobj\ just outside the adopted $Q<0.718$~au \ayloname{} boundary.
\par\smallskip\noindent\begin{minipage}{\columnwidth}
\makeatletter\def\@captype{table}\makeatother
\caption{JPL solution~\#2 for \scobj\ at the integration epoch.}\label{tab:orbit}
\centering
\begin{tabular}{lrr}
\hline\hline
Quantity & Value & Formal $1\sigma$\\\hline
$a$ (au) & 0.4957788889 & $3.8078\times10^{-7}$\\
$e$ & 0.4499795594 & $1.6645\times10^{-6}$\\
$q$ (au) & 0.2726885229 & $8.2154\times10^{-7}$\\
$Q$ (au) & 0.7188692548 & $1.0571\times10^{-6}$\\
$i$ (deg) & 16.77506435 & 0.00065822\\
$\Omega$ (deg) & 151.49518209 & 0.00047330\\
$\omega$ (deg) & 2.91886639 & 0.00085693\\
$M$ (deg) & 113.46368966 & 0.00196555\\
$t_{\rm p}$ (JD TDB) & 2461134.313066075 & 0.00067586\\\hline
\end{tabular}
\tablefoot{Heliocentric J2000 ecliptic elements at the covariance and integration epoch, JD~2461174.5~TDB (2026 May 14). Symbols denote semimajor axis $a$, eccentricity $e$, perihelion $q$, aphelion $Q$, inclination $i$, ascending-node longitude $\Omega$, argument of perihelion $\omega$, mean anomaly $M$ and perihelion time $t_{\rm p}$. Uncertainties propagate the full correlated covariance (Appendix~\ref{app:clones}).}
\end{minipage}
\par\smallskip

For $r-i$, we combine two exposure pairs measured with empirical point-spread functions (PSFs). We transform Pan-STARRS1 (PS1) reference stars directly to Cousins $R_C/I_C$ \citep{Tonry2012}, solve for exposure zero points and instrumental colour terms, and apply the \citet{Jordi2006} conversion to SDSS-equivalent AB. The $g-r$ and $i-z$ estimates use direct PS1-to-SDSS calibration. Table~\ref{tab:colours} gives the selected measurements. The interleaved filter sequence gives $r-i$ pairs separated by 1.129 and 4.603~min, much shorter than the rotation periods of several hours common among asteroids larger than approximately 200~m \citep{Warner2009}. For such periods, rotational brightness changes within each pair are expected to be small. The two $r-i$ estimates agree within their uncertainties (Table~\ref{tab:colours}).

\section{Results}\label{sec:colours}\label{sec:dynamics}
The SDSS-equivalent colours are $g-r=0.83_{-0.13}^{+0.21}$, $r-i=0.176_{-0.054}^{+0.052}$ and $i-z=-0.50_{-0.12}^{+0.11}$~mag. We integrate the 24 Bus--DeMeo class templates \citep{DeMeo2009,DeMeo2020} against the \citet{Willmer2018} solar spectrum and SDSS responses. The V template gives $\chi^2_\nu=1.116$ for three degrees of freedom, followed by R (2.726) and O (3.629). The comparison uses the formal colour uncertainties (Appendix~\ref{app:methods}). The \citet{Ivezic2001} diagnostic $a^*=0.89(g-r)+0.45(r-i)-0.57$ gives $0.24_{-0.12}^{+0.19}$~mag (Fig.~\ref{fig:astar}a).

Figure~\ref{fig:spectrum}b compares the composite reflectance with the class templates; Appendix~\ref{app:methods} defines its conversion and normalisation.

We checked the catalogue absolute magnitude against the calibrated Keck and FORS2 photometry. For an assumed $H$--$G$ phase-law parameter $G=0.15$, we obtain $H_V=18.80$~mag at Keck and 18.56~mag at the VLT. At $G=0.30$ the estimates become 19.11 and 18.86~mag. Our two epochs sample high phase angles of $89.4^\circ$ and $93.6^\circ$ (Appendix~\ref{app:methods}). We adopt the JPL value $H_V=18.83$~mag for the size estimates. With $H=H_V$,
$D=1329\,10^{-H/5}/\sqrt{p_V}$~km gives two conditional size estimates.
The NEOWISE V-type median $p_V=0.362$ \citep{Mainzer2011Taxonomy}
gives $D=0.379^{+0.027}_{-0.025}$~km. The model-predicted albedo estimate
$\widetilde p_V=0.222$ for \scobj\ \citep{Bolin2026Atira}
gives $D=0.483^{+0.034}_{-0.032}$~km. These intervals propagate the formal catalogue uncertainty of 0.15~mag at fixed albedo. The uncertainties in $H_V$ and hence $D$ are likely underestimated because the unknown phase function must be extrapolated from these high phase angles to zero phase. The model estimate uses the probabilities of three albedo
categories (Sect.~\ref{sec:conclusions}). Optically selected taxonomic
samples favour higher-albedo objects because they are brighter at a given
size. The V-type albedo scatter of 0.100 corresponds to
$D=0.335$--$0.445$~km at the central $H$, an albedo-sensitivity range rather than a confidence interval for this object.

The 5-Myr integrations in each direction use 10\,000 clones drawn from the correlated covariance of the adopted JPL~\#2 solution at JD~2461174.5~TDB. All residence and collision statistics refer to this ensemble (Appendix~\ref{app:clones}). REBOUND/MERCURIUS \citep{Rein2019} uses a two-day base timestep, the Sun, eight massive planets, a separate Moon and massless clones; the Newtonian model omits general-relativistic and nongravitational forces. Histories are sampled every 1000~yr (Appendices~\ref{app:clones} and \ref{app:collisions}). Trapezoidal integration of $n(Q<0.718~{\rm au})/10\,000$ gives mean cumulative residence of 231~kyr (46.1\%) during the past 500~kyr and 120~kyr (23.9\%) during the next 500~kyr. Removed clones contribute zero thereafter. Over the 100~kyr preceding the integration epoch, the ensemble-mean residence fractions are 30.2\% as \ayloname\ asteroids and 69.8\% as Atiras. Over the following 100~kyr, these fractions are 3.6\% and 96.0\%, respectively; 0.4\% of ensemble time follows collision removal. We classify bound orbits with $Q<0.718$~au as \ayloname\ and those with $0.718\leq Q<0.983$~au as Atira. Figure~\ref{fig:orbit} shows the recovery-day orbital geometry and the evolution of $q$ and $Q$.

Over $+5$~Myr, 1855 clones (18.55\%) undergo recorded planetary collisions: 377 with Mercury (3.77\%), 1456 with Venus (14.56\%), 22 with Earth (0.22\%) and none with Mars. By $+2$~Myr, 169 clones hit Mercury, 864 hit Venus and six hit Earth (1039 in total). A further 208, 592 and 16 collisions, respectively, occur between $+2$ and $+5$~Myr (816 in total). We remove clones at detected planetary surface overlaps in the forward integrations.

\section{Discussion and conclusions}\label{sec:conclusions}
The NEOMOD3 analysis of \citet{Bolin2026Atira} explicitly includes \scobj\ and assigns 46.2\%, 27.9\% and 21.3\% to delivery through the $\nu_6$, 3:1 and weak inner-main-belt resonances, respectively. Their sum, 95.4\%, favours delivery through inner-Main-Belt escape channels. Young families such as Datura demonstrate the production of sub-kilometre fragments in Main Belt collisions \citep{Vokrouhlicky2017}, supplying material that can enter these resonances. A basaltic fragment from a differentiated parent is consistent with the V-like colours. The Vesta family is a plausible reservoir, although other differentiated bodies may also have supplied inner-Main-Belt V-types \citep{Troianskyi2023}; the present data do not identify a parent family. Spectroscopy is needed to confirm the V-like classification indicated by our photometry \citep[e.g.,][]{Bolin2023Dinkinesh}.

For \scobj, the same population model predicts probabilities of $0.53\pm0.02$ for intermediate albedo ($0.1\leq p_V<0.3$), $0.37\pm0.02$ for high albedo ($p_V\geq0.3$) and $0.10\pm0.01$ for low albedo ($p_V<0.1$) \citep{Bolin2026Atira}. Using representative albedos of 0.05, 0.20 and 0.30 for these categories, we obtain the model-predicted albedo estimate $\widetilde p_V=0.05P_{\rm low}+0.20P_{\rm intermediate}+0.30P_{\rm high}=0.222$, where $P$ denotes each category probability. This prediction gives the second diameter estimate in Sect.~\ref{sec:colours}. The V-type median lies in the high-albedo category. Thermal observations could constrain the diameter and albedo.

At the integration epoch, $Q$ exceeds the \ayloname\ boundary by $8.7\times10^{-4}$~au, placing \scobj\ in the Atira class. Its clone ensemble crosses the inner-Venus boundary (Fig.~\ref{fig:qQ}b), connecting its current orbit with temporary \ayloname\ residence, as found for (594913) \ayloname\ \citep{Bolin2022Aylo}. These results indicate that material delivered from the inner Main Belt can move between the \ayloname\ and Atira populations.

\balance
\begin{acknowledgements}
Based on observations collected at the European Southern Observatory under ESO programme 116.2AP3.002 (PI: B.~T.~Bolin).
We thank Claudia Paladini for her support as Shift Coordinator for the VLT observations.
Some of the data presented herein were obtained at Keck Observatory, which is a private 501(c)3 non-profit organisation operated as a scientific partnership among the California Institute of Technology, the University of California, and the National Aeronautics and Space Administration. The Observatory was made possible by the generous financial support of the W. M. Keck Foundation. The raw Keck Observatory data presented herein, once public, can be accessed from the Keck Observatory Archive (KOA), which is operated by the W. M. Keck Observatory and the NASA Exoplanet Science Institute (NExScI), under contract with the National Aeronautics and Space Administration.
The authors wish to recognise and acknowledge the very significant cultural role and reverence that the summit of Maunakea has always had within the Native Hawaiian community. We are most fortunate to have the opportunity to conduct observations from this mountain.
The orbital integrations used resources of the Bura supercomputer at the Center for Advanced Computing and Modelling, University of Rijeka, Croatia.
The Keck observations were obtained under target-of-opportunity programme 2026B\_K450 (PI: L.~Y.~Alcorn). We thank M.~Lundquist and R.~Morris for observing support and the observing team of J.~X.~Prochaska and L.~Kahinga for accommodating the interruption.
D.F. conducted this research at the Jet Propulsion Laboratory, California Institute of Technology, under a contract with the National Aeronautics and Space Administration (80NM0018D0004).
\end{acknowledgements}
\paragraph{Data availability.} Tables~\ref{tab:colours} and \ref{tab:newastrometry} give the FORS2 colours and the Keck and FORS2 astrometry. The raw images are associated with ESO programme 116.2AP3.002 and Keck programme 2026B\_K450.
\bibliographystyle{aa}
\bibliography{references}
\FloatBarrier

\begin{appendix}
\onecolumn
\nobalance
\setlength{\dblfloatsep}{\SCsavedfloatsep}
\setlength{\dbltextfloatsep}{\SCsavedtextfloatsep}
\section{Initial orbit, covariance and clone ensemble}\label{app:clones}\label{app:orbitgeometry}
The clone ensemble uses archived JPL solution~\#2, dated 2026 October 12, based on 22 observations from 2025 September 27 to 2026 October 7, including T16 (375-d SBDB arc). Table~\ref{tab:orbit} gives elements at the native covariance and integration epoch, JD~2461174.5~TDB, distinct from the top-level SBDB epoch, JD~2461200.5. The covariance coordinates are $(e,q,t_{\rm p},\Omega,\omega,i)$. We derive $a=q/(1-e)$, $Q=a(1+e)$ and $M=\sqrt{GM_\odot/a^3}(t-t_{\rm p})$, where $GM_\odot$ is the solar gravitational parameter and $t$ is the epoch; $M$ is converted to degrees. Their uncertainties propagate the full covariance, including cross terms. Horizons propagates these elements to JD~2461320.709184634~TDB for Fig.~\ref{fig:orbit}a and to the observing midpoints for the geometry and residuals.

\begin{figure*}[h!]
\centering\includegraphics[width=158mm]{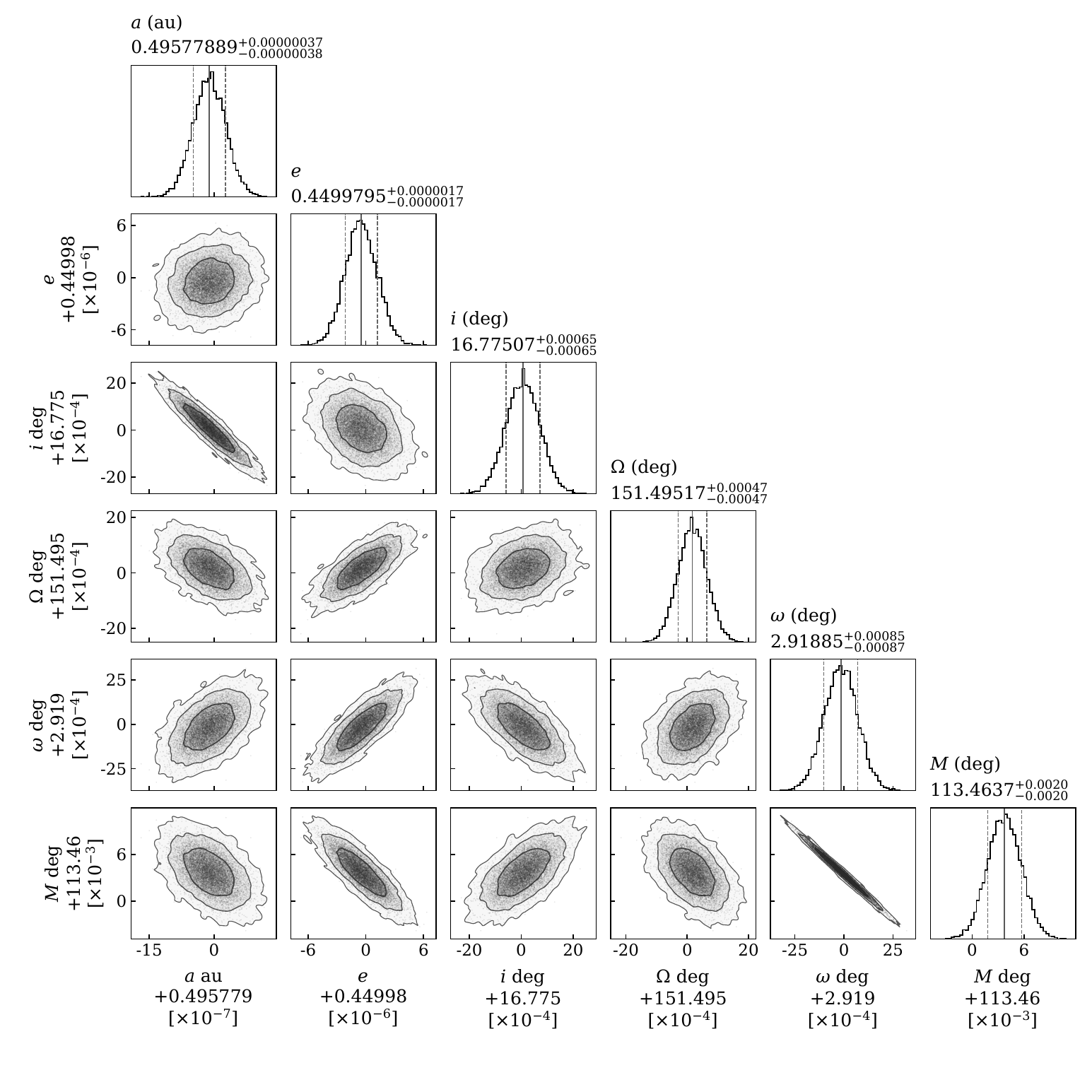}
\caption{Heliocentric J2000 ecliptic orbital elements of the 10\,000 initial JPL~\#2 clones at JD~2461174.5~TDB, ordered as $(a,e,i,\Omega,\omega,M)$. Diagonal histograms show marginal distributions; titles and solid lines give means, with dashed lines at the 16th--84th percentile bounds (central 68\% intervals). Translucent points mark individual clones above the shaded contours, which enclose approximately 68\%, 95\% and 99.7\% of the smoothed binned density, from inside out. Coordinates equal the printed offset plus the tick value times the indicated scale; $a$ is in au and angles are in degrees.}\label{fig:corner}
\end{figure*}

We reconstruct the symmetric covariance from its column-major upper-triangular representation in the native element order. Cholesky sampling of the correlation matrix, scaled by the native standard deviations, gives the element draws. We convert each draw to a heliocentric Cartesian state and add the Sun's barycentric state to initialise the integrations. Figure~\ref{fig:corner} shows heliocentric osculating $(a,e,i,\Omega,\omega,M)$ recovered from these states. Titles give arithmetic means and central 68\% intervals (16th--84th percentiles), which quantify the marginal orbital uncertainty of the clones.

For validation, we subtract the archived solar state and rotate the draws to the J2000 equator. The nominal position agrees with archived heliocentric Horizons~\#2 to 0.24~m. An independent Jacobian transformation gives equatorial position standard deviations of $(590,645,559)$~km along $(X,Y,Z)$, versus $(588,647,556)$~km for the draws. All six position and velocity standard deviations agree within 0.62\%; the maximum absolute correlation-coefficient difference is 0.0085.

\FloatBarrier
\clearpage\twocolumn
\section{Photometric calibration and taxonomic comparison}\label{app:methods}
\begin{table*}[t]
\caption{FORS2 colours in the SDSS-equivalent AB system.}\label{tab:colours}
\centering
\begin{tabular}{llll}
\hline\hline
Measurement / observing epoch & $g-r$ (mag) & $r-i$ (mag) & $i-z$ (mag)\\\hline
FORS2, 2026 Oct. 24--25 & $0.83_{-0.13}^{+0.21}$ & $0.176_{-0.054}^{+0.052}$ & $-0.50_{-0.12}^{+0.11}$\\
FORS2, first selected pair & -- & $0.178_{-0.065}^{+0.064}$ & --\\
FORS2, second selected pair & -- & $0.172_{-0.095}^{+0.093}$ & --\\\hline
\end{tabular}
\tablefoot{The selected $g-r$ measurement uses exposures $g5,13,17$ and $R2,10$; $i-z$ uses $I7,19$ and $z4,8$. The $r-i$ pairs are $R6/I7$ and $R14/I11$, separated by 1.129 and 4.603~min, respectively. Their intervals propagate target measurement noise, calibration covariance and Jordi coefficient uncertainties. Full-precision values are retained in the source data.}
\end{table*}

The 20-s FORS2 exposures began between 2026 October 24 at 23:43:55 and October 25 at 00:11:17~UTC. The topocentric observing geometry was calculated with Horizons at all 20 midpoint epochs by propagating the archived JPL~\#2 elements from the covariance epoch, using the Paranal observing site (309). The asteroid is on CHIP1. The total exposure times of the display stacks are 60, 40, 40 and 40~s in $g,R,I,z$, respectively. Registration in the world coordinate system (WCS) uses bilinear interpolation; each pixel requires unmasked contributions from every selected exposure.

We select 8544 distinct identified asteroids from the SDSS Moving Object Catalog (MOC1) with $2<a<3.5$~au, $0\leq e<0.4$, $0<\sigma_{g,r,i}\leq0.05$~mag and $0<\sigma_z\leq0.08$~mag, retaining the observation with the smallest summed $griz$ variance for each designation. Plot colours encode position in the colour--colour diagram. Template crosses and the Sun use solar-weighted SDSS colours. The adopted solar colours are $(0.460306,0.120955,0.020814)$~mag in $(g-r,r-i,i-z)$. The V mean predicts $(0.713291,0.175962,-0.325796)$~mag, so the template $i-z$ is 0.172~mag redder than the measured colour, or 1.63 times its upper formal uncertainty.

The reduced FORS2 primary arrays contain science data in analogue-to-digital units (ADU) and retain the sky background. Separate extensions supply the sky estimate, variance in ADU$^2$ and quality mask. Cleaned stamps contain measured data after subtraction of the fitted neighbours and sky background. Simultaneous CHIP1 and CHIP2 files belong to one exposure. The calibrated colour and reflectance comparisons follow earlier multi-filter studies of asteroids \citep{Bolin2020CD3,Bolin2023Trojans,Bolin2025KY26,Bolin2025YR4} and comets \citep{Bolin2025ThreeIDiscovery,Bolin2026ThreeIMonitoring,Bolin2026N2}.

With $x=(g-r)_{\rm PS1}$, we transform the reference-star magnitudes using
\begin{align}
R_C &= r_{\rm PS1}-0.137-0.108x-0.029x^2,\\
I_C &= i_{\rm PS1}-0.366-0.136x-0.018x^2
\end{align}
from \citet{Tonry2012}. The offsets already include the PS1-AB to Cousins-Vega system difference. After the instrumental colour-term solution, we apply
\begin{equation}
(r-i)_{\rm SDSS}=(1.007\pm0.005)(R_C-I_C)-(0.236\pm0.003)
\end{equation}
from \citet{Jordi2006}. We propagate shared calibration uncertainties into the generalised least-squares mean of the two pairs using a joint 3000-replicate star-block bootstrap and 50\,000 Monte Carlo draws of the target and calibration parameters. The pair weights are approximately 0.685842 and 0.314158. Jordi coefficient errors are treated as independent of one another and shared between pairs.

For the absolute-magnitude check, we transformed the photometry at each epoch separately to Johnson $V$. The nearly simultaneous Keck $g25$ and $R39$ give $g=20.94$ (AB) and $R_{\rm C}=19.90$~mag (Vega). Combining the $g-r$ and $r-R_{\rm C}$ relations of \citet{Jordi2006} gives $V-R_{\rm C}=(g-R_{\rm C}+0.051)/1.913=0.568$~mag, on the branch $V-R_{\rm C}\leq0.93$. This gives $V=20.47\pm0.07$~mag, propagating the shared instrumental colour correction. For FORS2 we applied the Cousins calibration to the two $R$ frames used for $g-r$, obtaining $R_{\rm C}=19.61$ and 19.66~mag. The stellar relation $g-r=1.646(V-R_{\rm C})-0.139$ gives $V-R_{\rm C}=0.586$~mag, with a colour-only uncertainty of $-0.081/+0.130$~mag. We reduced the resulting $V$ magnitudes to heliocentric and observer distances of 1~au and extrapolated to zero phase with the $H$--$G$ phase law \citep{Bowell1989HG} at each assumed $G$. We averaged the two FORS2 estimates.

The first and last retained $g$ measurements, separated by 17.18~min, differ by $0.349\pm0.114$~mag, with the uncertainty propagated from their statistical errors. The colours use the interleaved exposure selections in Table~\ref{tab:colours}.

For a reflectance template $R(\lambda)$, solar spectral flux density $F_\odot(\lambda)$ and total response $T_b(\lambda)$ in band $b$, the photon-weighted band reflectance is
\begin{equation}\label{eq:integration}
\langle R\rangle_b=\frac{\int R(\lambda)F_\odot(\lambda)T_b(\lambda)\lambda\,d\lambda}{\int F_\odot(\lambda)T_b(\lambda)\lambda\,d\lambda}.
\end{equation}
We obtain synthetic colours by adding the solar AB colour to $-2.5\log_{10}(\langle R\rangle_b/\langle R\rangle_c)$. The original template samples span 0.45--2.45~$\mu$m in 0.05-$\mu$m steps. Linear interpolation is used for numerical integration. We use constant $R(0.45~\mu\mathrm{m})$ below the short-wavelength limit when integrating the $g$ response. The $r,i,z$ responses lie within the template coverage.

We rank the templates using
\begin{equation}
\chi^2_\nu=\frac{1}{3}\sum_{j=1}^{3}\left[\frac{C_{j,\rm model}-C_{j,\rm obs}}{\sigma_{j,\rightarrow\rm model}}\right]^2.
\end{equation}
Here $C_j$ is the $j$th colour. We use its upper formal uncertainty when the model colour exceeds the measured value and its lower uncertainty otherwise. We compare three colour residuals with fixed templates and fit no continuous parameters, giving three nominal degrees of freedom.

For the composite display let $d_{gr}=(g-r)-(g-r)_\odot$, and similarly define $d_{ri}$ and $d_{iz}$. The $r$-relative ratios are
\begin{equation}
(u_g,u_r,u_i,u_z)=(10^{-0.4d_{gr}},1,10^{0.4d_{ri}},10^{0.4(d_{ri}+d_{iz})}).
\end{equation}
The plotted values $R_b=2u_b/(u_g+1)$ are $(0.833435,1.166565,1.227630,0.761627)$. The nominal $g$ and $r$ wavelengths (0.477 and 0.623~$\mu$m) bracket 0.55~$\mu$m symmetrically. Their mean reflectance therefore gives the linearly interpolated midpoint. Templates are normalised to their monochromatic reflectance at 0.55~$\mu$m. We use natural cubic splines to interpolate the plotted Bus--DeMeo spectra \citep{DeMeo2009}. Error bars propagate the asymmetric colour endpoints through this normalisation, combining same-direction log-reflectance excursions in quadrature.

\begin{table}[h!]
\caption{Three-colour comparison with 24 Bus--DeMeo templates.}\label{tab:taxonomy}
\centering
\begin{tabular}{lrlr}\hline\hline
Class & $\chi^2_\nu$ & Class & $\chi^2_\nu$\\\hline
V & 1.116 & Cg & 9.211\\
R & 2.726 & Xe & 9.322\\
O & 3.629 & B & 9.633\\
Sa & 3.782 & Xk & 10.025\\
Q & 4.195 & C & 10.141\\
Sr & 5.257 & Xc & 10.365\\
A & 5.294 & Cgh & 10.776\\
Sq & 5.568 & Cb & 11.128\\
S & 6.555 & T & 11.235\\
Sv & 6.963 & X & 11.247\\
K & 7.707 & Ch & 11.282\\
L & 9.013 & D & 12.660\\\hline
\end{tabular}
\tablefoot{Scores compare fixed templates with three degrees of freedom using the integration and weighting prescription above.}
\end{table}
\FloatBarrier

\clearpage\onecolumn
\section{Orbital evolution and collision diagnostics}\label{app:collisions}
\begin{figure*}[h!]
\centering\includegraphics[width=180mm]{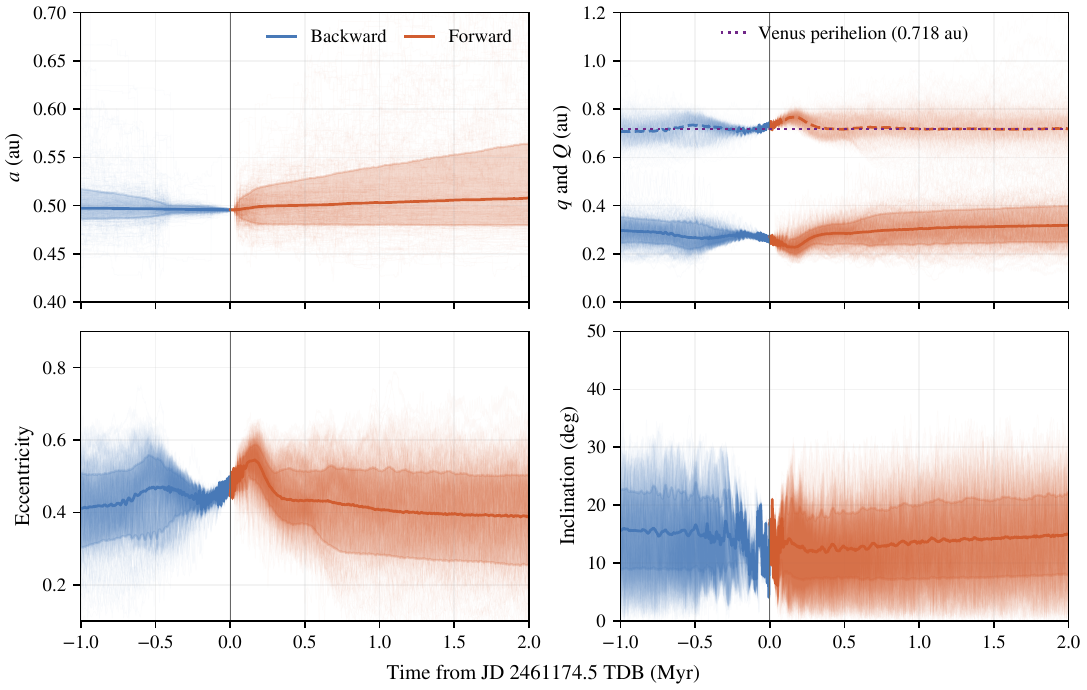}
\caption{Evolution of the JPL~\#2 ensemble from $-1$ to $+2$~Myr relative to JD~2461174.5~TDB. Panels show $a$, $q$ and $Q$ together, $e$, and $i$; blue and orange denote backward and forward integration. Faint lines show 400 clone histories at 1000-yr cadence. Solid curves and shading give the median and 16th--84th percentiles for $a,q,e,i$ over clones present at each epoch; dashed lines give median $Q$. The purple dotted line marks Venus's adopted perihelion distance (0.718~au). Summaries include trajectories outside the displayed vertical ranges. Collision counts through $+5$~Myr are reported in Sect.~\ref{sec:dynamics}.}\label{fig:evolution}
\end{figure*}

Figure~\ref{fig:evolution} shows 400 clone histories over $-1$ to $+2$~Myr at 1000-yr cadence. At $+2$~Myr, 8961 clones remain, and their median semimajor axis is 0.507852~au, with a central 68\% interval of 0.479915--0.564395~au. At $-1$~Myr, the corresponding median is 0.497231~au and the interval is 0.485707--0.517340~au. Of the initial 10\,000 clones, 44.62\% satisfy $Q<0.718$~au at $+2$~Myr and 57.88\% at $-1$~Myr. None satisfy this criterion at the integration epoch.

The integrations use REBOUND~5.2.2 with \texttt{safe\_mode=1}; the clone ensemble was generated with random seed 20261012. Removal is restricted to forward surface overlaps with Mercury, Venus, Earth and Mars; backward collisions, solar-collision removal and an ejection cutoff are disabled. At $+5$~Myr, 8145 clones remain, including two on unbound osculating orbits, and 3644 satisfy $Q<0.718$~au. At $-5$~Myr, 27 clones have unbound osculating orbits.

\FloatBarrier

\begin{figure*}[h!]
\centering\includegraphics[width=180mm]{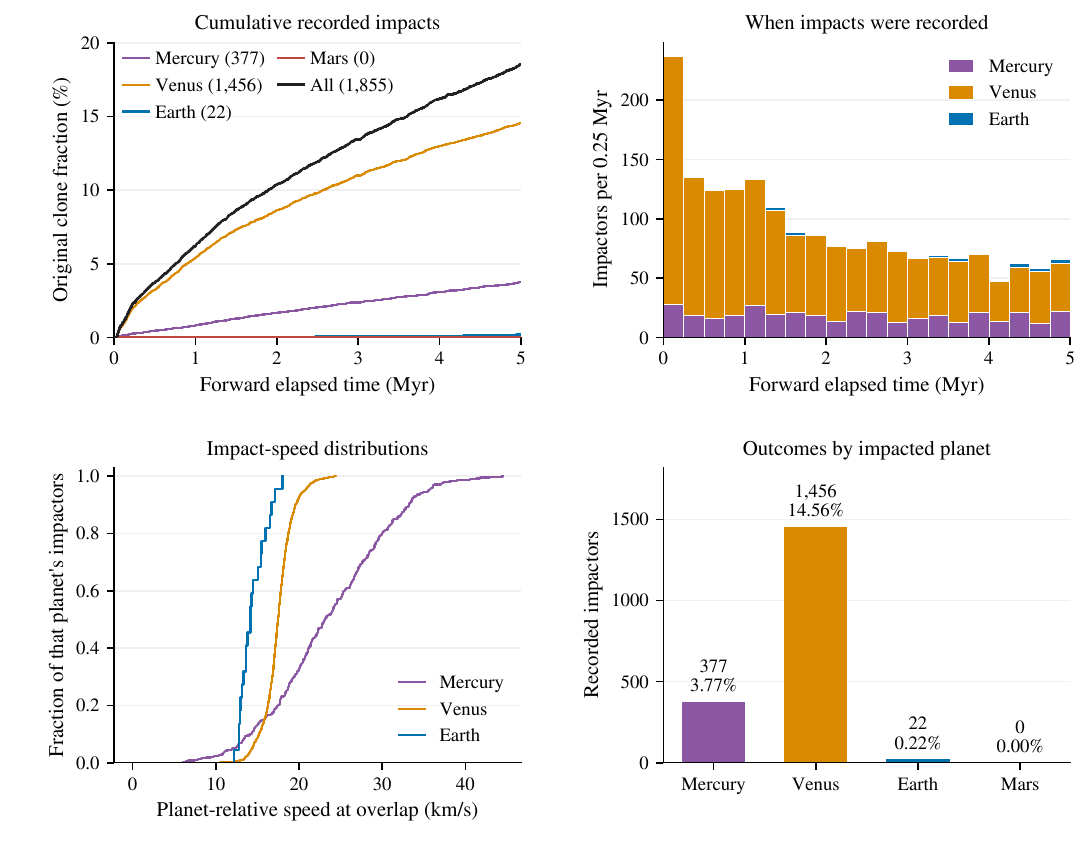}
\caption{Forward collision diagnostics for the JPL~\#2 ensemble through $+5$~Myr. Upper left: cumulative collision fractions relative to the original 10\,000 clones, by planet and in total. Upper right: stacked collision counts in 0.25-Myr bins. Lower left: empirical cumulative distributions of planet-relative speed at the detected surface overlap. Lower right: collision counts and percentages of the initial ensemble, by planet. Colours identify the planets; the black curve in the upper-left panel gives the total. Events are the finite-step surface overlaps defined in Sect.~\ref{sec:dynamics}.}\label{fig:collisionsfull}
\end{figure*}

Event times are recorded when finite-step surface overlaps are detected (Fig.~\ref{fig:collisionsfull}). For Mercury, Venus and Earth, respectively, the median recorded times are 2.287, 1.489 and 3.682~Myr, and the median relative speeds are 23.086, 17.493 and 14.173~km~s$^{-1}$. Speeds are calculated from the relative Cartesian event states. Finite-step overlap detection can miss grazing events.

\FloatBarrier

\clearpage\onecolumn
\section{Keck and FORS2 astrometry}\label{app:newastrometry}
Table~\ref{tab:newastrometry} gives the four Keck/LRIS recovery positions, the four retained FORS2 positions and their offsets from solution~\#2. For LRIS, we fitted elliptical Gaussian profiles with a sloping background to determine centroids, and calibrated cubic tangent-plane solutions against Gaia DR3 star positions propagated from epoch 2016.0 using their proper motions. The $g$-camera solution includes an empirical step near detector row 512. Reference-star leave-one-out residuals span 0.013--0.034~arcsec per coordinate. The adopted coordinate errors are listed in Table~\ref{tab:newastrometry}.

LRIS times are the midpoints of the intervals between the shutter events recorded in \texttt{DATE\_BEG} and \texttt{DATE\_END}; all four integrations last 30~s. FORS2 midpoints use $\texttt{MJD-OBS}+\texttt{EXPTIME}/(2\times86400)$, with exposure time in seconds. The error budget includes a 1-s timing term for both instruments. It is included in the tabulated Keck coordinate uncertainties and recorded separately for FORS2, where it corresponds to approximately 0.026~arcsec in $\alpha\cos\delta$ and 0.018~arcsec in $\delta$.

\begin{table*}[h!]
\caption{Keck and FORS2 astrometry and observed-minus-computed offsets from JPL solution~\#2.}\label{tab:newastrometry}
\centering
\small\setlength{\tabcolsep}{4pt}\begin{tabular}{llrrrrrrr}
\hline\hline
Frame/filter & Midpoint UTC & Exposure & $\alpha$ (ICRS) & $\delta$ (ICRS) & $\sigma_{\alpha\cos\delta}$ & $\sigma_\delta$ & $\Delta\alpha\cos\delta$ & $\Delta\delta$\\
 & & (s) & (deg) & (deg) & (arcsec) & (arcsec) & (arcsec) & (arcsec)\\\hline
LRIS R38 & 2026-10-07 05:00:04.370 & 30 & 237.29086119 & -19.16352131 & 0.307 & 0.305 & $+0.095$ & $-0.060$\\
LRIS g24 & 2026-10-07 05:00:09.000 & 30 & 237.29089700 & -19.16353456 & 0.109 & 0.104 & $+0.024$ & $+0.003$\\
LRIS R39 & 2026-10-07 05:01:09.917 & 30 & 237.29162750 & -19.16393511 & 0.109 & 0.103 & $-0.030$ & $+0.019$\\
LRIS g25 & 2026-10-07 05:01:18.000 & 30 & 237.29172076 & -19.16399417 & 0.112 & 0.106 & $-0.050$ & $0.000$\\
FORS2 R2 & 2026-10-24 23:45:44.214 & 20 & 253.398518459 & -28.154325825 & 0.10832 & 0.10642 & $-0.160$ & $+0.029$\\
FORS2 R6 & 2026-10-24 23:51:29.488 & 20 & 253.401353914 & -28.156002060 & 0.11085 & 0.10987 & $-0.026$ & $+0.094$\\
FORS2 R10 & 2026-10-24 23:57:12.708 & 20 & 253.404126306 & -28.157685227 & 0.10307 & 0.10380 & $-0.045$ & $+0.096$\\
FORS2 R18 & 2026-10-25 00:08:39.710 & 20 & 253.409616618 & -28.161073515 & 0.11509 & 0.11271 & $-0.287$ & $+0.027$\\
\hline
\end{tabular}
\tablefoot{MPC codes are T16 (Keck) and 309 (FORS2). LRIS uses native $g/R$ filters; FORS2 uses $R_{\rm SPECIAL+76}$. Coordinates and times reproduce the Keck PSV and selected FORS2 ADES XML precision. The tabulated adopted uncertainties use the timing treatment described above.}
\end{table*}

For FORS2, we fitted empirical PSFs for the target and neighbouring stars with a local planar sky in the full reduced CHIP1 images, using their variance and masks. Gaia DR3 stars supplied an affine correction to the existing simple-imaging-polynomial (SIP) solution. The coordinate uncertainties combine centroid, fitting-region and plate-solution terms with a 0.10-arcsec systematic term.

\begin{figure*}[h!]
\centering\includegraphics[width=178mm]{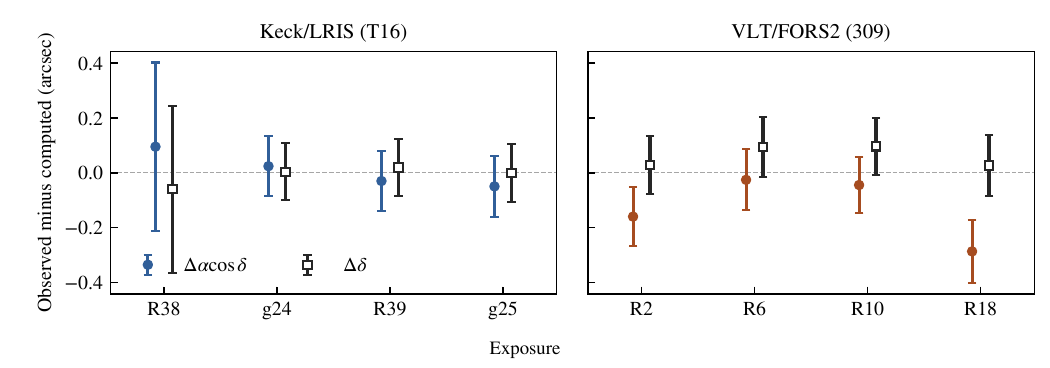}
\caption{Observed minus computed topocentric astrometric offsets of \scobj\ relative to JPL solution~\#2, by exposure. Filled circles show $(\alpha_{\rm obs}-\alpha_{\rm calc})\cos\delta$ and open squares show $\delta_{\rm obs}-\delta_{\rm calc}$. Error bars give the measurement uncertainties in Table~\ref{tab:newastrometry}. The FORS2 timing term is recorded separately as described above.}\label{fig:newresiduals}
\end{figure*}

We compared all eight measurements with the topocentric astrometric ICRF positions calculated by propagating the archived JPL solution~\#2 with Horizons at the UTC midpoints for T16 and 309 (Fig.~\ref{fig:newresiduals}). We do not use the W57 measurements flagged as rejected in Davide Farnocchia's supplied JPL residual report. The Keck radial residuals range from 0.02 to 0.11~arcsec. The four FORS2 positions lie 0.03--0.29~arcsec west of the prediction, with declination offsets smaller than 0.10~arcsec in magnitude.

\FloatBarrier

\end{appendix}
\end{document}